# Generation of Nonlocal Spaces from Moyal Brackets and Penrose Lattices by way of example

Takeo Miura

e-mail : t_miura@msg.biglobe.ne.jp

Advanced Institute for Mathematical Science,Ltd.

1-24-4,Motojyuku,HigashiMatuyama,Saitama,355-0063 ,Japan

January 31, 2007

## Abstract

In this paper we consider a connection between the Moyal formulation of quantum mechanics and noncommutative lattices on nonlocal spaces. The Moyal bracket generates SU(N) Lie algebras, and representations of these algebras correspond to Bratteli diagrams on Hilbert spaces of $\mathbb{C}$*-algebras, and also representations of noncommutative lattices on nonlocal spaces correspond to the same Bratteli diagrams. We will show that the quantization thus generates nonlocal spaces by the intermediation of Bratteli diagrams. The above connection (intermediation) are investigated in details in two definite cases, one is one-dimensional lattice QCD SU(N) fields on the unit circle, the other is Penrose lattice SU($q_N$) fields on the Penrose lattice.

## Ⅰ. Introduction

A foundation of quantum mechanics is quantum conditions. As one of quantum conditions, there is the Moyal bracket on phase space. An infinite dimensional SU(∞) algebra and finite dimensional SU(N) subalgebras are capable of being constructed by functions satisfying the Moyal bracket. A matrix ring representation of a finite dimensional SU(N) algebra sequence in a Hilbert space is found by Fairle et al [1]. In the following, we show two kinds of notions (1),(2), and they shall be equivalent.

（1）In this paper, we show a result that is a sequence of the representation of embedding maps of matrix rings corresponding to Bratteli diagrams. As its concrete examples, we have two examples, one is a square lattice sequence appeared in the lattice QCD and the finite subalgebras of it are noncommutative, but it becomes to be commutative at N → ∞ limit, moreover remains noncommutative under the restriction where N is any finite number which satisfying the codition $\hbar \leq 2\pi/N$, and the other is noncommutative SU($q_N$) algebras on the Penrose tiling lattice, which remain noncommutative at the limit N → ∞. By the way, different Bratteli diagrams take different embedding maps, the reason is attributed to differences of Primitive ideals.

（2） In this paper, we propose a notion of a nonlocal space. In physical detection where to distinguish events arise in certain small area, Balachandran,et al[4], Landi[5] introduced a partial order topology into spaces of the area, and the spaces become non-Housedorff spaces, therefore they identify any two points in certain sets.

A noncommutative lattice can be constructed on this space, and there are Bratteli diagrams which correspond to primitive ideals on the $\mathbb{C}$*-algebra. We will regard not to be distinguishable as Nonlocality, this nonlocality may be essential of physical space.

Now, we noticed a connection which connect the notion of the above(1)to the notion of the (2) by the matrices $U_n$, $V_n$ and their Bratteli diagrams. By this intermediation (connection) we can introduce the Nonlocality of physical space from the Moyal bracket condition, in the category of quantum mechanics.

Summary of each chapter is described as follows. In chapter Ⅱ, the Moyal bracket is expressed with finite dimensional SU(N) algebras. In chapter Ⅲ,we show that finite dimensional SU(N) algebra sequence corresponds to matrix ring sequence. In chapter Ⅳ,we show that the matrix ring sequence corresponds to Bratteli diagram. In chapter Ⅴ,the generation of the nonlocal space from Bratteli diagram is described. In chapter Ⅵ, we offer summary and prospect.

## Ⅱ. Finite dimensional SU(N) algebras of Moyal bracket

The sine ,or Moyal, bracket {{f,g}}, namely the extension of the Poisson brackets{f,g} to statistical distributions on phase-space, introduced by Weyl[6] and Moyal[7], and explored by several authors[9] in alternative formulation of quantum mechanics, understood as a deformation of the algebra of classical observables. It is generalized convolution which reduced to the Poisson bracket as $\hbar$, replaced by 2k in our context, is taken to zero.

$$\{\{f,g\}\}=(-r/4\pi^2 k^2)\int d p'\, d p''\, d x'\, d x''\, f(x', p')g(x'', p'')\sin((1/k)[p(x'-x'')+x(p''-p')+p'x''-p''x']) \qquad (1)$$

The argument of the sine is

$$\frac{1}{k}\begin{pmatrix}1 & p & x \\ 1 & p' & x' \\ 1 & p'' & x''\end{pmatrix}=\frac{1}{k}\int pdq$$

i.e. 2/k times the area of the equilateral phase-space triangle with vertices at (x,p), $(x', p')$, and $(x'', p'')$.

(1) Finite dimensional SU(N) algebras

What meaning is the right hand side of the Moyal bracket equation (1)? .We will describe it according to the analysis in references[1]. The Fourier-series of the functions f(x,p) and

g(x,p) on a phase-spaces are, respectively,

$$f(x,p) = - \sum_{m_1,m_2} F(m_1,m_2)\exp[i(m_1 x + m_2 p)], \qquad (2\text{-}1)$$

$$g(x,p) = - \sum_{n_1,n_2} G(n_1,n_2)\exp[i(n_1 x + n_2 p)], \qquad (2\text{-}2)$$

As a concrete example of the phase space, we choose a two-dimensional phase-space, $0 \leq x, p \leq 2\pi$. We can also choose the case of a genus-one closed surface with the topology of a torus $T^2 = S^1 \times S^1$ and we take the two variables $\sigma_1$ and $\sigma_2$ to parameterize the angle of two cycles in $S^1 \times S^1 : (\sigma_1, \sigma_2) \in [0, 2\pi] \times [0, 2\pi]$, this choice for the two angle is possible because that Deformations of the Lie algebra of a $\mathbb{C}^\infty$ function on the symplectic manifold, defined by the Moyal bracket, can generate the Heisenberg Equation[9].

For easy to understand the meaning of the following procedure, we select arbitrarily only one component from each Fourier expansions of (2-1) and (2-2), respectively, and put these components into (1). These components, for examples, are as follows,

$$f_{\hat{m}} = -\exp\{i(m_1 x + m_2 p)\},$$
$$g_{\hat{n}} = -\exp[i(n_1 x + n_2 p)],$$

Then, it is straightforward to check (1).
we obtain

$$\{\{f,g\}\}_{\hat{m},\hat{n}} = 2i\sin(k(\hat{m} \times \hat{n}))K_{\hat{m}+\hat{n}}, \qquad (3)$$

where

$$K_{\hat{m}+\hat{n}} = \frac{i}{2k}\exp[i(m_1+n_1)x + i(m_2+n_2)p] \qquad (4)$$

Now, we take x and p to be local (commuting) coordinates for the phase space, and take f and g to be differentiable functions of them, and also we can define a basis-independent realization for the generators of the algebras as follows,

$$L_f = (\partial f/\partial x)\partial/\partial p - (\partial f/\partial p)\partial/\partial x$$

so that

$$[L_f, L_g] = L_{\{f,g\}},$$
$$[L_f, g] = \{f, g\},$$

where $\quad \{f,g\} \equiv (\partial f/\partial x)\partial g/\partial p - (\partial f/\partial p)\partial g/\partial x$

This is a Poisson bracket of classical phase-space.

The generators $L_f$ thus transforms (x, p) to (x−∂f/∂p, p+∂f/∂x). Infinitesimally, this is a canonical transformation which preserves the phase-space area element dxdp.

The Fourier expansions of the $L_f$ are

$$L_f = \sum_{m_1,m_2} F(m_1,m_2) L(m_1,m_2),$$

$$L(m_1,m_2) = \exp[i(m_1 x + m_2 p)](i m_1 \partial/\partial p - i m_2 \partial/\partial x)$$

Further evaluate

$$L(m_1,m_2) \equiv L_{\hat{m}},$$
$$[L_{\hat{m}}, L_{\hat{n}}] = (\hat{m} \times \hat{n}) L_{\hat{m}+\hat{n}} \qquad (5)$$

These algebras are the SU(N)Lie algebras, and for the limit N→∞, these reduces to the SU(∞)Lie algebra.

Now, what means the right hand side of (3)?. We show it below.

We define the $K_f$

$$K_f = \frac{1}{2ik} f(x + ik\partial/\partial p, p - ik\partial/\partial x),$$

the following is the first approximate term,

$$K_f = \frac{1}{2}(\partial f/\partial x \cdot \partial/\partial p - \partial f/\partial p \cdot \partial/\partial x)$$

this is the same to $L_f$.

the term which chose in the Fourier expansions of the above $K_f$ is the following,

$$K_{(m_1,m_2)} = \frac{i}{2k} \exp(i m_1 x + k m_2 \partial/\partial x + i m_2 p - k m_1 \partial/\partial p)$$

$$= \frac{i}{2k} \exp(i m_1 x + i m_2 p) \exp(k m_2 \partial/\partial x - k m_1 \partial/\partial p)$$

For the small $k$ limit, this reduces to

$$= \frac{i}{2} \exp(i m_1 x + i m_2 p)(\frac{1}{k} + m_2 \partial/\partial x - m_1 \partial/\partial p)$$

and for $\frac{1}{k} \gg m_2 \partial/\partial x - m_1 \partial/\partial p$, we get

$$K_{\hat{m}} = K_{(m,m)} = \frac{i}{k} \exp(i m_1 x + i m_2 p)$$

We have thus found the following algebras

$$[K_{\hat{m}}, K_{\hat{n}}] = (\frac{i}{2})^2 [\exp(i m_1 p + i m_2 x)\exp(i n_1 p + i n_2 x) - \exp(i n_1 p + i n_2 x)\exp(i m_1 p + i m_2 x)]$$

At this time, we consider the coordinates x and p to be the operators X and P, and we define the commutator [X,P]=i, thus, we can introduce the quantization. This method is due to H.Weyl[6]. As the coordinates have changed to the operators, we can use the Baker−Campbell−Hausdorff's formula.
We then obtain ,by using this formula

$$[\mathrm{K}_{\hat{m}},\mathrm{K}_{\hat{n}}] = \frac{1}{k}\sin[k(\hat{m}\times\hat{n})]\mathrm{K}_{\hat{m}+\hat{n}} \qquad (6)$$

the above (6) has the same algebraic structure of $L_{\hat{m}}$ (5), at the limit k→0.
From (3),(6), we obtain

$$\{\{f,g\}\}_{\hat{m},\hat{n}} = 2ik[\mathrm{K}_{\hat{m}},\mathrm{K}_{\hat{n}}] = i\hbar[\mathrm{K}_{\hat{m}},\mathrm{K}_{\hat{n}}]$$

Thus, SU(N)Lie algebras are generated as the right hand side of the above equation.
as a consequence, the bracket $[\mathrm{K}_{\hat{m}},\mathrm{K}_{\hat{n}}]$ is introduced by means of the quantization. The reason for this introduction is the following : Deformations of the Lie algebra of $\mathbb{C}^\infty$ functions, defined by the Poisson bracket, generalized the Moyal bracket Deformations of the algebra of $\mathbb{C}^\infty$ functions, defined by ordinary multiplication, give rise to noncommutative, associative algebras, isomorphic to the operator algebras of quantum theory [9]. The Deformation quantization means that the Moyal bracket is equivalent to the ordinary quantum conditions.

### Ⅲ. Matrix ring representations of finite dimensional SU(N) algebras

We describe similarly to the analysis in reference [1]. The structure constants of (6) remain invariant under the modular transformation of the index vectors.

This induces the automorphism $\mathrm{K}_m \to \mathrm{K}_{m'}$. Let us now focus on the cyclotomic family: the one for which $k = 2\pi/N$, for integer $N > 2$. In this family, there is an additional $\mathbb{Z}\times\mathbb{Z}$ algebra isomorphism $\mathrm{K}_{(m_1,m_2)} \to \mathrm{K}_{(m_1,m_2)+(Nt,Nq)}$ for arbitrary integers $t$ and $q$. Since the structure constants $\sin(2\pi/N)(m_1 n_2 - n_1 m_2)$ are only sensitive to the modulo-$N$ values of indices, the two-dimensional lattice separates into $N\times N$ cell. The fundamental $N\times N$ cell contains $N^2$ index points, but the operator $\mathrm{K}_{(0,0)}$, like its lattice translations $\mathrm{K}_{N(t,q)}$, factors out of the algebra. Thus the fundamental cell involves only $N^2 - 1$ generators, and there are no more structure constants than those occurring in this cell. In consequence, the infinite dimensional centerless cyclotomic algebras, with the $K_{N(t,q)}'$ s factored out, posses the finite dimensional ideal of "lattice average" operators 𝕂.

This $(N^2 - 1)$-dimensional ideal specifies, in fact, a basis for SU(N). For simplicity, consider odd $N'$ first. A basis for SU($N$) algebras, odd $N$, may be built from two unitary unimodular matrices:

$$g \equiv \begin{pmatrix} 1 & 0 & \cdots & 0 \\ 0 & \omega & \cdots & 0 \\ \vdots & \vdots & \ddots & \vdots \\ 0 & 0 & \cdots & \omega^{N-1} \end{pmatrix}, \qquad (7)$$

$$h \equiv \begin{pmatrix} 0 & 1 & \cdots & 0 \\ \vdots & \vdots & \ddots & \vdots \\ 0 & 0 & \cdots & 1 \\ 1 & 0 & \cdots & 0 \end{pmatrix},$$

$g^N = h^N = I$,

$\omega = \exp(4\pi i/N)$,

hg=$\omega$gh, (8)

The complete set of unitary unimodular $N \times N$ matrices

$$J_{(m_1, m_2)} \equiv \omega^{m_1 m_2 /2} g^{m_1} h^{m_2},$$

where

$$J^{+}_{(m_1, m_2)} = J_{(-m_1, -m_2)}$$

$\mathrm{Tr} J_{(m_1, m_2)} = 0$ except for $m_1 = m_2 = 0$ mod N,

suffice to span the algebra of SU($N$). They close under multiplication to just one such, by virtue of (8):

$$J_m J_n = \omega^{m \times n /2} J_{m+n}$$

They consequently satisfy the the algebra

$$[J_m, J_n] = -2i(\sin(2\pi/N)(\hat{m} \times \hat{n})) J_{m+n} \qquad (9)$$

let divide the both side of (9) by $(k)^2 = (2\pi/N)^2$ 、 and replace $(1/k)J_m$ by $K_m$,

$$[K_m, K_n] = -(2i/k)(\sin(k(\hat{m} \times \hat{n}))) K_{m+n} \qquad (10)$$

(10) is exactly the preceding (6), thus we obtain the representation of $K_m$ by matrix ring.

For even $N$, the fundamental matrices in (7) are not unimodular, as their determinant may now be -1 as well. One might choose to modify them to

$$g \equiv \sqrt{\omega} \begin{pmatrix} 1 & 0 & \cdots & 0 \\ 0 & \omega & \cdots & 0 \\ \vdots & \vdots & \ddots & \vdots \\ 0 & 0 & 0 & \omega^{N-1} \end{pmatrix},$$

$$h \equiv \begin{pmatrix} 0 & 1 & \cdots & 0 \\ \vdots & \ddots & \ddots & \vdots \\ 0 & 0 & \cdots & 1 \\ -1 & 0 & \cdots & 0 \end{pmatrix}$$

$g^N = h^N = -1$, they again obey (8).

### IV. Matrix ring sequences and its Bratteli diagrams

A basis for SU(N) algebras may be build from two unitary unimodular matrices, In short, these are generators. These generators are related to Bratteli diagrams. The exact description of these relations is not described in this article while referring to Pimusner and Voiculescu[8],Landi[5] for a more detailed account. We confine ourselves only to describing two concrete examples, one is two-dimensional Penrose tiling and the other is two point partial order set(pos).

（1） In the case of Penrose tiling [8],[2]

Let θ be an irrational number and $A_\theta$ the corresponding irrational rotation $\mathbb{C}$*-$A_\theta$ algebra ( i.e. the cross-product of the continuous functions on the unit circle by the automorphism corresponding to the rotation of angle $2\pi\theta$ ). The irrational rotation algebra is generated by two unitaries $u, v \in A_\theta$ such that $u^*vu = e^{2\pi i \theta} v$ and every pair $U, V$ of unitaries in some unital $\mathbb{C}$*-algebra $B$, satisfying $U^*VU = e^{2\pi i \vartheta} V$ determines a *-monomorphism ψ: $A_\theta \to B$ ,, such that $\psi(v) = V$, $\psi(u) = U$.

Let $[a_1, a_2, ..., a_n, ...]$ be the continued fraction expansion of θ. The $n'$ th convergent $\frac{p_n}{q_n}$ equals $[a_1, a_2, ..., a_n]$ and $p_n, q_n$ may be computed recurrently by the formulae :

$$p_n = a_n p_{n-1} + p_{n-2} \quad (n \geq 2) \text{ with } p_0=0, \ p_1=1$$
$$q_n = a_n q_{n-1} + q_{n-2} \quad (n \geq 2) \text{ with } q_0=1, \ q_1=a_1. \tag{11}$$

For each $n \in \mathbb{N}$ let $M_{q_n}$ denote the $q_n \times q_n$ matrices acting on the Hilbert space $\mathbb{C}^{q_n}$.

The canonical basis in $\mathbb{C}^{q_n}$ will be denoted by $e(n; j), 1 \leq j \leq q_n$. The $AF$-algebra into which we shall imbed $A_\theta$ is the inductive limit

$$\to A_{n-1} \xrightarrow{\rho_n} A_n \to$$

of the finite-dimensional $\mathbb{C}$*-algebras $A_n = M_{q_n} \oplus M_{q_{n-1}}$, the imbeddings $\rho_n : A_{n-1} \to A_n$ being defined by

$$\rho_n(x \oplus y) = W_n(x \oplus ... \oplus x \oplus y)W_n^* \oplus x$$

where $x \in M_{q_{n-1}}, y \in M_{q_{n-2}}$ and

$$W_n : \mathbb{C}^{q_{n-1}} \oplus \ldots \oplus \mathbb{C}^{q_{n-1}} \oplus \mathbb{C}^{q_{n-2}} \to \mathbb{C}^{q_n}$$

is a unitary operator we shall define below. Remark that (11) makes the definition of $W_n$ possible.

The operator $W_n$ (n ≥ 6) is of the form:

$$W_n(\xi_1 \oplus \ldots \oplus \xi_{a_n} \oplus \eta) =$$
$$= W_n^{(k)}\xi_1 + \ldots + W_n^{(a_n)}\xi_{a_n} + \tilde{W}_n \eta$$

where

$$W_n^{(k)} : \mathbb{C}^{q_{n-1}} \to \mathbb{C}^{q_n}, \quad (1 \leq k \leq a_n)$$

$$\tilde{W}_n : \mathbb{C}^{q_{n-2}} \to \mathbb{C}^{q_n}.$$

For each $n \geq 4$ we shall consider the operators $U_n, V_n \in M_{q_n}$ defined by:

$$U_n e(n; j) = e(n; j+1) \quad \text{when } 1 \leq j < q_n$$
$$= e(n;1) \quad \text{when } j = q_n$$

and

$$V_n e(n; j) = \zeta_n^j e(n; j) \quad \text{where } \zeta_n = \exp(2\pi i \frac{p_n}{q_n})$$

The convergence conditions (see [8]) are

$$\| \rho_n(U_{n-1} \oplus U_{n-2}) - U_n \oplus U_{n-1} \| \leq \frac{300\pi}{q_{n-2}},$$

$$\| \rho_n(V_{n-1} \oplus V_{n-2}) - V_n \oplus V_{n-1} \| \leq \frac{42\pi}{q_{n-1}} + \frac{7\pi}{q_{n-2}}.$$

thus, $A_n$ satisfys the inductive limit.

We emphasize particularly that the matrices $U_n, V_n \in M_{q_n}$ are identified with h,g $\in$ SU($q_n$)

and $W_n, \tilde{W}_n$ are embedding operators $A_{n-1} \to A_n$.

In the case of Penrose tiling, we take θ to be τ, this is called *Golden Ratio*.

h,g in the preceding chapter are correspond to $U_n, V_n$ (the n*th* matrix) in Penrose tiling.

$$V_n U_n = \exp(2\pi i \cdot p_n / q_n) U_n V_n$$

$p_n / q_n = \tau_n$

At the time $n \to \infty$, $\tau_n \to \tau$.

n is the n*th* index of the matrix sequence, the matrix of SU($q_n$) is $q_n \times q_n$ matrix.

$$V_n = \begin{pmatrix} 1 & 0 & \cdots & 0 \\ 0 & \omega & 0 & \vdots \\ \vdots & \cdots & \ddots & \vdots \\ 0 & \cdots & \cdots & \omega^{q_{n-1}} \end{pmatrix}$$

$$U_n = \begin{pmatrix} 0 & 1 & \vdots & 0 \\ \vdots & \ddots & \ddots & \vdots \\ \vdots & \vdots & \ddots & 1 \\ 1 & 0 & \cdots & 0 \end{pmatrix},$$

$\omega = \exp(2\pi i p_n / q_n)$

The construction of basis is the following

$$J_m = J_{(m_1, m_2)} = \omega^{m_1 m_2 / 2} U_m^{m_1} V_m^{m_2}$$

$$J_m J_n = \omega^{m \times n / 2} J_{m+n},$$

obtain SU($q_n$), and $\quad [J_m, J_n] = -2i(\sin((\pi p_n / q_n) m \times n)) J_{m+n}$ (12)

At the limit n→∞、ω→exp($2\pi i \tau$), $V_n \to V$, $U_n \to U$, and VU=$e^{2\pi i \tau}$ UV、VU remains not commutative.

In the case of Penrose lattice, (12) remains a noncommutative algebra even at the limit n→∞, but a cyclotomic lattice is different, (9) changes to commutative algebra. The cyclotomic lattice corresponds to one dimensional variation of a square lattice, and the Lattice QCD is the theory of a square lattice, these results show that the Lattice QCD becomes a commutative and ordinarily continuous spaces, at the limit n→∞.

The algebra $A_{PT}$ of The Penrose Tiling is the AF algebra (Approximately finite Algebra) which has the kernel{0}, which is the only primitive ideal. The construction of its K-theory (see [11],[5]) is rather straightforward .

At each level of inclusion( in Bratteli diagram ), the algebra is given by

$$A_n = M_{d_n}(\mathbb{C}) \oplus M_{d'_n}(\mathbb{C}), \quad n \geq 1,$$

with inclusion.

The Penrose inclusion map is described with two kinds of isosceles trangle's inclusion.

$$I_n : A_n \hookrightarrow A_{n+1}, \quad \begin{Bmatrix} A & 0 \\ 0 & B \end{Bmatrix} \mapsto \begin{Bmatrix} A & 0 & 0 \\ 0 & B & 0 \\ 0 & 0 & A \end{Bmatrix}, \quad (13)$$

for any $A \in M_{d_n}(\mathbb{C}), B \in M_{d'_n}(\mathbb{C})$. This gives, for the number of dimensions, the following recursive relations,

$$d_{n+1} = d_n + d'_n,$$

$$d'_{n+1} = d_n, \qquad n \geq 1, \ d_1 = d'_1 = 1,$$

In the introduction, we described the intermediation of (1) and (2). We describe the notion in details [11].

Let denote by $Q_n(A)$ for the set of idempotents in the matrix of $M_n(A)$, and $SU(n) \subset GL_n(A)$ for the group of invertible elements in $M_n(A)$.

There are canonical identifications

$$\varphi: \ M_n(A) \hookrightarrow M_{n+1}(A), \qquad m \mapsto \varphi(m) = \begin{pmatrix} m & 0 \\ 0 & 0 \end{pmatrix},$$

and likewise

$$\varphi: \ Q_n(A) \hookrightarrow Q_{n+1}(A), \qquad e \mapsto e \oplus 0 = \begin{Bmatrix} e & 0 \\ 0 & 0 \end{Bmatrix}.$$

For invertible matrices we identify

$$\psi: \ GL_n(A) \hookrightarrow GL_{n+1}(A), \qquad v \mapsto \begin{pmatrix} v & 0 \\ 0 & 1 \end{pmatrix}.$$

These $U_n, V_n \in SU(n) \subset GL(n)$ are Generators, and introduced in the previous and this chapter.

We will give the definition of projector.

Given a unital $\mathbb{C}^*$-algebra $A$, we denote by $M_n(A) \simeq M_n(\mathbb{C}) \otimes_C A$ the $\mathbb{C}^*$-algebra of n×n matrices with entries in $A$. Two projectors $p,q \in M_n(A)$ are said to be equivalent if there exists a matrix $u \in M_n(A)$ such that $p = u^*u$ and $q = uu^*$. In order to be able to 'add' equivalence classes of projectors, one considers all finite matrix algebras over $A$ at the same time by considering $M_\infty(A)$ which is the non complete *-algebra obtained as the inductive limit of finite matrices,

$$M_\infty(A) = \bigcup_{n=1}^{\infty} M_n(A),$$

$$\varphi: \ M_n(A) \to M_{n+1}(A), \qquad a \mapsto \varphi(a) = \begin{Bmatrix} a & 0 \\ 0 & 0 \end{Bmatrix}.$$

Now, two projectors $p,q \in M_\infty(A)$ are said to be equivalent, $p \sim q$, when there exists a $u \in M_\infty(A)$ such that $p = u^*u$ and $q = uu^*$. The set $V(A)$ of equivalence classes $\{\cdot\}$ is made into an abelian semigroup by defining an *addition*,

$$[p] + [q] =: \left[ \begin{Bmatrix} p & 0 \\ 0 & q \end{Bmatrix} \right], \qquad \forall [p], [q] \in V(A).$$

The additive identity is just $0 =: [0]$. p and q are projectors, and they are expressed with $U_n$ and $V_n$ in this Chapter and Chapter Ⅲ.

$$p = U_n{}^* U_n, \quad p^* = U_n U_n{}^*, \quad p = p^* = p^2,$$
$$q = V_n{}^* V_n, \quad q^* = V_n V_n{}^*, \quad q = q^* = q^2,$$

In conclusion, $U_n$, $V_n$ are identified with h,g, and they make the intermediation described in the chapter. I. Penrose tiling shows a geometrical point of view on τ(irrational number), on the other hand the continued fractional expansion is the number theoretical point of view on τ. Pimsner and Voiculescu[8] obtained the unified view of the both views, on the $\mathbb{C}^*$-algebra.

(2) In the case of two point poset {p,q} [4]

We shall assume that algebra $A$ has a unit and $A_n + \mathbb{C} I$ is a finite dimensional $\mathbb{C}^*$-subalgebra of $A$. Let $\mathcal{H}$ be an infinite dimensional(separable) Hilbert space. The algebra

$$A = \mathcal{K}(\mathcal{H}) + \mathbb{C} I_H, \tag{14}$$

with $\mathcal{K}(\mathcal{H})$ the algebra of compact operators, is an AF-algebra([3],[4],[5]). The approximating algebras are given by

$$A_n = M_n(\mathbb{C}) \oplus \mathbb{C}, \quad n>0,$$

with embedding

$$M_n(\mathbb{C}) \oplus \mathbb{C} \ni (\Lambda,\lambda) \mapsto \left( \begin{Bmatrix} \Lambda & 0 \\ 0 & \lambda \end{Bmatrix}, \lambda \right) \in M_{n+1}(\mathbb{C}) \oplus \mathbb{C},$$

The norm closure of $\bigcup_n A_n$ is the following,

$$\overline{\cup_n A_n} \simeq \mathcal{K}(\mathcal{H}) + \mathbb{C} I = A.$$

The algebras (14) has only two irreducible representations,

$$\pi_1 : A \to \mathcal{B}(\mathcal{H}), \quad a = (k + \lambda_1 I_H) \mapsto \pi_1(a) = a,$$
$$\pi_2 : A \to \mathcal{B}(\mathbb{C}), \quad a = (k + \lambda_2 I_H) \mapsto \pi_2(a) = k,$$

with $\lambda_1, \lambda_2 \in \mathbb{C}$, and $k \in \mathcal{K}(\mathcal{H})$; the corresponding kernels being

$$\mathfrak{I}_1 = \ker(\pi_1) = \{0\},$$
$$\mathfrak{I}_2 = \ker(\pi_2) = \mathcal{K}(\mathcal{H}).$$

The partial order given by the inclusions $\mathfrak{I}_1 \subset \mathfrak{I}_2$ produces the two point poset.

(V) From Bratteli diagrams to Nonlocal spaces

The Bratteli diagram is the diagram of a finite dimensional matrices representation of Primitive ideals, in Hilbert spaces [4],[5]. When an algebra $A$ is the inductive limit of the finite dimensional algebra $A_n$, the set of all compact $k \in \mathcal{K}(\mathcal{H})$ in a Hilbert space is a $\mathbb{C}^*$-algebra. $A$ is a two-sided ideal in the $\mathbb{C}^*$-algebra $\mathcal{B}$ of all bounded operators. These finite dimensional spaces are the orthogonal component of null space of Hilbert spaces.

Now, according to the Gel'fand-Naimark theorem, an algebra $C(\mathcal{M})$ of continuous functions which is isomorphic to $\mathbb{C}^*$-algebra, be able to be constructed on a structure space $\mathcal{M}$. $\mathcal{M}$ is the space of equivalence class of irreducible representation(IRR), and $\mathfrak{I}_1=\ker(\pi_1), \mathfrak{I}_2=\ker(\pi_2)$ are the structure space, the topology of the structure space is an order topology by $\mathfrak{I}_1 \subset \mathfrak{I}_2$ (Jacobson Topology)、therefore, the space with partial order topology is a non-Hausdorff space, which is a nonlocal spase in this case.

Let M be a continuous topological space, and we assume that the sets $\{O_\lambda\}$ cover M 、and M consists of finite numbers(N) of $O_\lambda$ with partial order topology. We denote it by $\mathbb{P}_N(\mathcal{M})$, $\mathbb{P}_N(\mathcal{M})$ has $T_0$ topology space. The number of sets $O_\lambda$ is finite when $\mathcal{M}$ is compact, so that $\mathbb{P}(\mathcal{M})$ can be approximate to M by the covering of the finite set $\{O_\lambda\}$.

As an equivalence class is denoted by $\sim$, $\mathbb{P}(\mathcal{M}) \equiv \mathcal{M}/\sim$ is the quotient topology with equivalence class. $x \sim y$ means $x \in O_\lambda \Leftrightarrow y \in O_\lambda$ for every $O_\lambda$, and any two points $x, y \in \mathcal{M}$, if they coexist in certain set $O_\lambda$, can not be distinguished, then we regard them some extended existence(spread) and define this property as Nonlocality.

We propose then an Nonlocal topological space. The following description is quoted from references[4],[5]：「Experiments are never so accurate that they can detect events associated with points of M, rather they only detect events as occuring in certain sets $O_\lambda$. It is therefore natural to identify any two points x, y of M if they can never be separated or distinguished by the sets $O_\lambda$」. The above authors have discussed a noncommutative lattice with a finite approximation. In this article we regard a spread of events(matter) to be Nonlocal, the identification of two points cause noncommutativity to Quantum Mechanics.

### (VI) Summary and prospect

Several kinds of noncommutative QCD and noncommutative gauge field have been researching [10], and among those there is a field theory on the noncommutative space in which coordinates are directly made operations and $[x_\mu, x_\nu]=i\theta_{\mu\nu}$ is assumed from the beginning. $\theta_{\mu\nu}$ is generated under a background field. This theory is supposed as on the analogy of the gravitational field theory.

In this paper, it seems as if $\tau$ is considered as $\theta_{\mu\nu}$. We showed that nonlocality of spaces is generated by the quantization of fields. Nonlocality depends on ways of some finite coverings over the compact spaces by open sets. The deformation quantization may be regarded as one of theories which represent concretely nonlocality of spaces.

.A coming subject shall be a research for some kinds of Nonlocal Geometry and then,

approaches to this subject from the research of Penrose lattices (lattices with irrational number ratio) may supply some significant results.

References

[1]　D.B.Fairle, P.Fletcher and C.K.Zachos, Phys.Lett B. 218,2,(1989)203;
　　D.B.Fairlie and C.K.Zachos, Phys.Lett B, 224,1,2,(1989)101;
　　D.B.Fairlie,P.Fletcher and C.K.Zachos, J.Math.Phys.31.5.(1990)1088.

[2] T.Miura, talk on 「Noncommutativity of Penrose lattice」 at The 4th workshop of Quasi Periodic Tiling and Related Topics, at RIMS, Kyoto University, December, 2004;
　　T.Miura, Bulletin Of The Society For Science On Form. 19.1(2004)123.

[3] O.Bratteli, J.Functional Analysis.16.(1974)192.

[4] A.P.Balachandran,G.Bimonte,E.Ercolessi,G.Landi,F.Lizzi,G.Sparano,
　　P.Teotonio-Sobrinho, J.Geom.Phys.18(1996)163.

[5]G.Landi,Lecture Notes in Physics:Monograph m51, (Springer,Berlin,1997).

[6] H.Weyl, The Theory of Groups and Quantum Mechanics,(Dover,NewYork,1931).

[7] J.Moyal, Proc.Camb.Phil.Soc. 45(1949)99.

[8] M.Pimusner and D.Voiculescu, J.Operator Theory, 4(1980)201.

[9]F.Bayen,M.Flato,C.Fronsdal,A.Lichnerowicz,andD.Sternheimer,
　　Ann.Phys.(NY)111(1978)61,111;
　　H.Omori,Y.Maeda ,and A.Yoshioka., Adv.Math.85(1991)224;
　　B.V.Fedosov, J.Differential Geom, 40(1994)213.

[10] M.Chaichian, A.Demichev and P.Prošnajder, Nucl.Phys. B 567(2000),360;
　　J.Trampetic and J.Wess (Eds), Part IV, Noncommutative Field Theories, in 「 Particle Pysics in the New Millenium」 Lecture Notes in Physics,vol.616(2003),Springer,Berlin.

[11]José M.Garcia-Bondia, Joseph C.Várilly and Héctor Figeroa, Elements of Noncommutative Geometry, (Birkhäuser,2001).